\documentclass[aps,pra,10pt,twocolumn,showpacs,superscriptaddress,footnoteinbib]{revtex4}
\usepackage{amsmath,float,amsthm}
\usepackage{latexsym}
\usepackage{amssymb,dsfont}
\usepackage{color}
\usepackage{mathtools}
\usepackage{graphicx}
\usepackage[outdir=./]{epstopdf}
\usepackage{graphics}
\usepackage{soul}
\usepackage[breaklinks]{hyperref}

\newtheorem{theorem}{Theorem}
\begin{document}

\title{Self-testing of quantum states using symmetric local hidden state model}
\author{Debasis Mondal}
\thanks{cqtdem@nus.edu.sg} 
\affiliation{Centre for Quantum Technologies, National University of Singapore, 3 Science Drive 2, Singapore 117543}
\author{Dagomir Kaszlikowski}
\thanks{phykd@nus.edu.sg} 
\affiliation{Centre for Quantum Technologies, National University of Singapore, 3 Science Drive 2, Singapore 117543}
\affiliation{Department of Physics, National University of Singapore, 2 Science Drive 3, 117542 Singapore, Singapore}
\date{\today}


\begin{abstract}
We introduce a symmetric local hidden state $(slhs)$ model in a scenario, where two spacially separated parties receive quantum states from an unknown source. We derive an inequality based on the model. A completely new form of nonlocality emerges from the resource theoretic point of view. The inequality singles out a larger set of quantum correlated states in the higher dimensional scenarios $(d> 2 $ X $2)$ than what is predicted by the existing $lhs$ model, opening a new front for the experimentalists to test the accuracy of the prediction. We propose an experiment to show the experimental violation of the inequality in the two qubit scenario and perform the experiment on the IBM quantum computer. However, the experimental method adopted for the two-qubit scenario does not naturally generalize in the higher dimensional scenarios and leaves the experimental verification of the claim open. We also show that the maximal violation of the inequality can be used to self-test the Bell state and measurement bases, leading to complete device-independence.
\end{abstract}


\maketitle
\section{Introduction} In the last few decades, we have explored different avenues to understand the difference between the classical and the quantum world. Quantum correlation, for instance, have been of great interest for its quantum advantages over the classical in communication, cryptography and computation. For a bipartite system, quantum correlation depends on the protocol---following which the correlation is generated and the fact whether we trust the parties or not. A trustless model is where parties do not need to trust each other, for example, the local hidden variable $(lhv)$ model \cite{bell, jfclauser}. Entanglement, on the other hand, is a measure of the quantum correlation between two trusted parties. A seemingly new addition to the list is the local hidden state $(lhs)$ model \cite{wittmann,lhsintro, sjjones}, where only one of the parties is trusted.

A quantum state is shared by two parties--- Alice and Bob. The statement that the state of Alice is correlated to the state of Bob but not the state of Bob with that of Alice, sounds like: Alice is married to Bob but Bob is not \cite{nguyen,uola}. The violation of the present description of the local hidden state $(lhs)$ model \cite{lhsintro,sjjones, wittmann} sometimes predicts exactly that. There are states for which Alice can steer Bob but Bob cannot steer Alice. 

Although, the steerability of a quantum state has been shown to  be a resource \cite{resource2, resource3} and strongly believed to be a signature of quantum correlations, it cannot be considered as a measure of quantum correlations. This is due to the fact that quantum correlations like its classical counter part must be symmetric with respect to parties. Is it then possible to find an alternative and improved local hidden state model for which such scenarios do not arise? 

The situation arises primarily because of the game considered in the description of local hidden state model \cite{lhsintro, sjjones}. We consider two parties--- Alice and Bob. Alice prepares a bipartite state. A part of which she sends to Bob and keeps the other part with her. Alice claims, she can steer his state and his state is entangled with her. Bob does not trust Alice. Bob comes up with a strategy to verify her claim by constructing the steering inequality based on the $lhs$ model.

In the existing $lhs$ model \cite{lhsintro,sjjones, uola,wittmann,nguyen,resource2,resource3}, Bob assumes that the states, he is receiving from Alice, have single system description and Alice can in principle prepare such states from an ensemble of hidden states  $\{p(\lambda), \rho_{\lambda}\}$ such that
\begin{equation}
\rho_{a|\Pi}=\sum_{\lambda}p(\lambda)p(a|\Pi,\lambda)\rho_{\lambda},
\label{lhs}
\end{equation}
where $p(a|\Pi, \lambda)$ is Alice's stochastic map to convince Bob and $\lambda$ is a hidden variable such that $\sum_{\lambda}p(\lambda)=1$, $\Pi$ is an observable in which Bob asked Alice to perform measurements and $a$ is the outcome. Violation of a steering inequality based on the $lhs$ model implies the existence of $EPR$ non-locality.
 
 As it can be observed, the game is asymmetric by construction. A symmetric scenario would be, where Alice $(A)$ and Bob $(B)$ both receives quantum states from an unknown source $(S)$. They do not trust the source. Therefore, they would like to come up with an effective strategy to verify whether they have $EPR$-correlations between their particles or not.
 
 One way to represent the symmetric local hidden state $(slhs)$ model  is of course to express the bipartite state by local hidden states, i.e.,
 \begin{equation}
 \rho_{AB}=\sum_{\lambda}p(\lambda)\rho_{\lambda}^{A}\otimes\rho_{\lambda}^{B},
 \end{equation}
 where $\{p(\lambda), \rho_{\lambda}^{A}\otimes\rho_{\lambda}^{B}\}$ is an ensemble of hidden states like before. Alice and Bob can construct an inequality based on this model and the level of their trust. 
 
 Assumption in the above model is that any bipartite state must be expressible in terms of local hidden states. However, this is nothing but a separable state. Any state, which cannot be expressed by the above model is nothing but an entangled state and to witness an entangled state in the most effective way, Alice and Bob need to trust each-other. 
 
 To better understand how the trust is enforced in the measures of entanglement, it is noted that the entanglement measures depend on the elements of the form of $\langle i^{a}i^{b}|\rho_{AB}|i^{a}i^{b}\rangle$, where $ a, b\in \{0, 1\} \text{ for two qubit states}$. Therefore, measurements and outcomes need to be communicated to each-other to measure such quantities and the entanglement depends on the exact values of measurement outcomes. This now raises a question: is it possible to witness quantum nonlocality based on the local hidden state model in a trust-less manner? 
 
The present description of the $lhs$ model is asymmetric and as a result, only leads to the semi-device independence. It naturally creates a curiosity why there should not be a device-independent or trustless protocol to observe EPR nonlocality just like $lhv$ model? Moreover, the $lhs$ model is stronger than the $lhv$ model. Therefore, less quantum states have local hidden state description than local hidden variable description and more states violate the $lhs$ model than the $lhv$ model \cite{nguyen}. In this sense, nonlocality based on the violation of $lhs$ model is more accurate description of the nonlocality \cite{jones,pati} and a new symmetric, trustless description of such nonlocality should be better suited for device independent $QKD$ protocols than the existing protocols based on the Bell nonlocality and the EPR steering \cite{bennet,device,branciard}.

 It turns out that there is indeed another way to represent the $slhs$ model.  We use the $slhs$ model to self-test the state and the measurement bases, making it only the second, yet better alternative to the Bell inequality for device independent protocols and trust-less entanglement witness.

\section{Symmetric local hidden state model}  In the $lhv$ model \cite{bell, jfclauser}, we consider the joint probability distribution. We use the principle of locality and determinism to write the joint distribution in terms of local probability distributions and hidden variables. The locality in Bell's formalism does not imply EPR locality. Non-locality in a different form may still exist. It is just that the non-locality is not reflected in its local probability distributions.

Here, we assume that both Alice and Bob is receiving quantum states. However, quantum states provide more information than just probability distributions. One can also extract information about certain quantities, which have no classical counterparts unlike probability distributions. We consider a global or joint property with no classical counterpart and using the principle of locality and hidden states, express in terms of the local property of the hidden states. In particular, we consider the transition amplitude or probability to be the property, which has no classical counter part. With these assumptions, all the off-diagonal elements of a bipartite density matrix must be expressible by some unknown local hidden states $\rho_{A}^{\lambda}$ and $\rho_{B}^{\lambda}$ generated by the unknown source with a distribution $p(\lambda)$. This is probably the most general version of the $slhs$ model one could write. In a simple mathematical term, this implies,
  \begin{equation}
\langle i^{a}j^{b}|\rho_{AB}|i^{a'}j^{b'}\rangle=\sum_{\lambda}p(\lambda)\langle i^{a}|\rho^{\lambda}_{A}|i^{a'}\rangle\langle j^{b}|\rho^{\lambda}_{B}|j^{b'}\rangle
\label{strongerslhs1}
\end{equation}
for any arbitrary bipartite state $\rho_{AB}$, $\forall |{i}^{a(a')}\rangle$ and $|{j}^{b(b')}\rangle$. Here, $i$, $j$ are the labels for bases and $a$, $a'$, $b$ and $b'$ are the labels for vectors in the particular basis. In this section, we consider transition probability to express a relatively weaker version of our hidden state model in the following form,
 \begin{align}
 p\bigg{(}|{i}^{a-a'}\rangle^{A},|{j}^{b-b'}\rangle^{B}\bigg{)}=\sum_{\lambda}p(\lambda)p\bigg{(}|{i}^{a-a'}\rangle^{A}\bigg{|}\lambda\bigg{)}\nonumber\\p\bigg{(}|{j}^{b-b'}\rangle^{B}\bigg{|}\lambda\bigg{)}.
 \label{slhsmodel1}
 \end{align}
  Here, $p({i}^{a-a'})$ denotes the transition probability of the state from the $a$ Eigenvector to the $a'$ Eigenvector of the observable $i$ on Alice's side and $p(|{i}^{a-a'}\rangle)=|\langle {i}^{a}|\rho|{i}^{a'}\rangle|$. Note however that $p(|{i}^{a-a'}\rangle)$ does not really have properties of a probability distribution function. One can indeed show that the Eq. (\ref{strongerslhs1}) implies the Eq. (\ref{slhsmodel1}) (see the supplemental material \cite{supple}).

A clear distinction of the model from the local hidden variable $(lhv)$ model is that the $lhv$ model with dichotomic observables can explain the probability distributions generated by two classical coins. However, the same distribution cannot be described by the $slhs$ model of two qubit systems. The expression in Eq. (\ref{strongerslhs1}) or in Eq.  (\ref{slhsmodel1}) is an attempt to describe a (property of a) bipartite system in terms of that of the local hidden states. 
 
\section{ SLHS inequality}In this section, we derive an inequality based on the model for a two-qubit states. We consider the joint transition probability of Alice and Bob in the Eigen bases of the observables $i$ and $j$, $p(|{i}^{a-a'}\rangle^{A},|{j}^{a-a'}\rangle^{B})$ and express it in terms of the $slhs$ model as
 
 \begin{align}
 &\sum_{\lambda}p(\lambda)p\bigg{(}|{i}^{a-a'}\rangle^{A}|\lambda\bigg{)}p\bigg{(}|{j}^{a-a'}\rangle^{B}|\lambda\bigg{)}\nonumber\\ &\leq   \sqrt{\sum_{\lambda'}p(\lambda')p^{2}\bigg{(}|i^{a-a'}\rangle^{A}\bigg{|}\lambda'\bigg{)}\sum_{\lambda}p(\lambda)p^{2}\bigg{(}|j^{a-a'}\rangle^{B}\bigg{|}\lambda\bigg{)}}\nonumber\\& = \sqrt{\overline{p^{2}\bigg{(}|i^{a-a'}\rangle^{A}\bigg{)}}. \overline{p^{2}\bigg{(}|j^{a-a'}\rangle^{B}\bigg{)}}}\leq\frac{1}{4}\nonumber\\  &\implies  \underset{a\neq a'}{\sum_{a, a',}}p(|{i}^{a-a'}\rangle^{A},|{j}^{a-a'}\rangle^{B})\leq\frac{1}{2},
 \label{slhsineq}
 \end{align}

where in the first inequality, we use the Cauchy-Schwarz inequality, the second inequality comes from the fact that for a probability distribution $p(x)$ over a real random variable $x$ and a positive function of the random variable $f(x)$, if the the function is bounded from above by $l$, the $n^{th}$-moment of $f(x)$ is bounded by $l^n$. In this case, the function of the random variable $\lambda$ is $p\bigg{(}|i^{a-b}\rangle^{A}\bigg{|}\lambda\bigg{)}$ and is bounded by the maximal value of the transition probability of a local state can possibly take. 
For a two-qubit scenario, the dimension of the local state is two and the maximum transition probability between two arbitrary orthonormal states labeled by $a$, $b$ such that $a\neq b$ turns out to be $\frac{1}{2}$. Thus, the bound becomes $\frac{1}{4}$. The inequality can also be extended for a general $d\otimes d$ bipartite systems as shown later in Eq. (\ref{slhsineq5}). 

It is well known that Bell inequality cannot be violated by generating random measurement statistics locally. This makes the inequality trustless. It should be noted here that the violation of the inequality in Eq. (\ref{slhsineq}) ensures the existence of the nonlocality in the state and this inequality as well cannot be violated by generating random local measurement outcomes making it an another alternative to the Bell inequality in terms of trustless entanglement detection.

 \begin{figure}
\includegraphics[scale=0.6]{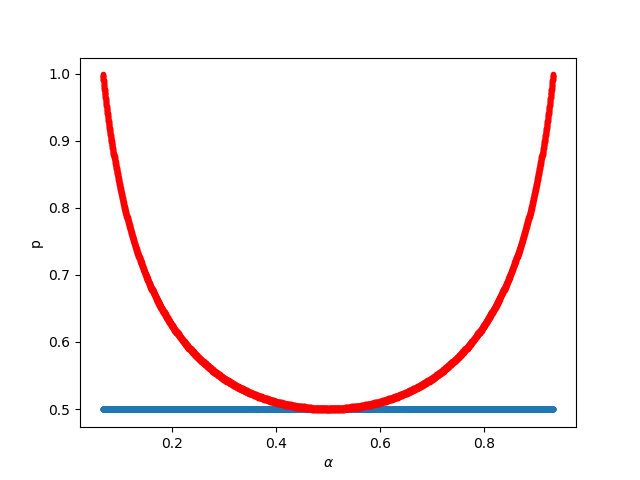}
\caption{We plot (in red) the left hand side of the Eq. (\ref{slhsineq}) with respect to the parameters for the Werner state $\rho_{w}^{\phi+}=p|\phi_{+}^{\alpha}\rangle\langle\phi_{+}^{\alpha}|+\frac{1-p}{4}I_{4}$, where $|\phi_{+}^{\alpha}\rangle=\sqrt{\alpha}|00\rangle+\sqrt{1-\alpha}|11\rangle$, when both Alice and Bob measures the transition probabilities in the $\sigma_3$ basis. The straight line parallel to the $\alpha$-axis represents the bound $p>\frac{1}{2}$, for which the state is nonlocal.}
\label{figurewerner}
\end{figure} 

As it can be observed from the fig. (\ref{figurewerner}), quantum states indeed violate the inequality. We will leave now the question on how Alice and Bob are going to measure the quantity on the left hand side of the inequality in Eq. (\ref{slhsineq}) for the next section and instead, we focus on to ask a more deeper question: what would happen if one considered quantum coherence instead of quantum transition probabilities? One would expect to get similar violation of a bound based on quantum coherence. More so due to the fact that quantum steering and quantum coherence have a deep connection \cite{deba1,deba2,deba3,deba4}. However, it is surprisingly not the case. One can follow similar steps to come up with a bound, which turns out to be a trivial bound with no violation by the bipartite entangled states.

{\it Free will.---} One can verify that the same Werner state as considered for the plot in fig. (\ref{figurewerner}), will not show violation for a similar inequality as
\begin{equation}
\underset{a\neq a'}{\sum_{a, a'}}p(|{i}^{a'-a}\rangle^{A},|{j}^{a-a'}\rangle^{B})\leq\frac{1}{2}.
\label{slhsineq2}
\end{equation}
The inequality in Eq. (\ref{slhsineq2}) is violated by yet another Werner state $\rho_{w}^{\psi+}=p|\psi_{+}^{\alpha}\rangle\langle\psi_{+}^{\alpha}|+\frac{1-p}{4}I_{4}$, where $|\psi_{+}^{\alpha}\rangle=\sqrt{\alpha}|01\rangle+\sqrt{1-\alpha}|10\rangle$.

One can derive another inequality based on these two inequalities in Eq. (\ref{slhsineq}) and (\ref{slhsineq2}) just by summing both the inequalities. The new inequality is although not tight, to show the violation, one does not need to invoke the assumption of free will. Both the Werner states for $p>\frac{1}{\sqrt{2}}$, violates the inequality as given below,
\begin{equation}
\underset{a\neq a', b\neq b'}{\sum_{a, a', b, b'=0}^1}p(|{i}^{a'-a}\rangle^{A},|{j}^{b-b'}\rangle^{B})\leq 1.
\label{slhsineq3}
\end{equation}
For a $d\otimes d$ bipartite system, an equivalent inequality corresponding to the inequality in Eq. (\ref{slhsineq2}) will be
\begin{equation}
\underset{a\neq a'}{\sum_{a, a'=0}^{d-1}}p(|{i}^{a'-a}\rangle^{A},|{j}^{a-a'}\rangle^{B})\leq\frac{d-1}{d}.
\label{slhsineq5}
\end{equation}
whereas, an equivalent inequality to that given in Eq. (\ref{slhsineq3}) will be
\begin{equation}
\underset{a\neq a', b\neq b'}{\sum_{a, a', b, b'=0}^{d-1}}p(|{i}^{a'-a}\rangle^{A},|{j}^{b-b'}\rangle^{B})\leq\bigg{(}d-1\bigg{)}^2,
\label{slhsineq4}
\end{equation}
The inequality in Eq. (\ref{slhsineq5}) is tighter than that in Eq. (\ref{slhsineq4}).
The two inequalities above do not follow directly from the two-qubit scenario. For a detailed analysis, see supplemental material \cite{supple}.

Let us now focus on to compare the inequalities with that of the existing bounds. For example, it is well-known that a generalized isotropic state belonging to the Hilbert space $\mathcal{H}_{d}\otimes\mathcal{H}_{d}$ of the form of 

\begin{equation}
\rho_{AB}=(1-p)\frac{I_{d^2}}{d^2}+\frac{p}{d}\sum_{i,j}|ii\rangle\langle jj|
\end{equation}

is steerable for $p>\frac{\sum_{r=2}^{d}\frac{1}{r}}{d-1}$ and entangled for $p>\frac{1}{d+1}$ \cite{nguyen}. A similar bound on $p$ can be derived from our inequalities and it turns out to be $p>\frac{1}{d}$.  Therefore, although our model predicts the same bound as the existing LHS model for a two-qubit Werner state, in the higher dimensions, our model predicts tighter bounds on $p$ than the existing model and opens a new avenue to test the new boundary of quantum correlations.

\section{ Experimental realization} 
\begin{figure}
\includegraphics[scale=0.58]{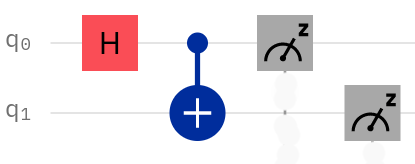}
\caption{
We provide a schematic diagram to generate entangled states and perform measurements in different bases. Here, $+$ denotes the $CNOT$ gate, $H$ stands for the Hadamard gate and $z$ stands for the measurement detector. To verify the quantum violation of the inequality in Eq. (\ref{slhsineq}), one does not need to measure the overlaps directly. Instead, one can use the identities as given in equations in Eq. (\ref{equivalentof5}) and Eq. (\ref{equivalentof6}) in the two qubit scenario.
}
\label{fig2}
\end{figure}

So far, $EPR$ steering has been observed in several
experiments \cite{wittmann,saunders,eberle,smith,bennet1,tischler,ksun,yxiao}. 
 We propose an experimental setup to show the violation of our inequality in this section. Before we proceed further on how to realize an experiment for quantum violation of the inequality just like the theoretical violation, let us first focus on to understand what does this quantity on the left hand side of Eq. (\ref{slhsineq}) represent? One could write any transition probability  of the state $\rho$ from  $|i^a\rangle$ to $|i^b\rangle$ as $p(|i^{a-b}\rangle)=|\langle i^a|\rho| i^b\rangle|$, where $i$ is the label depicting the eigenbasis of an observable and $a$ and $b$ are labels for the eigenvectors in the basis. For $\sigma_3$ observable, it turns out to be
\begin{equation}
p(|z^{0-1}\rangle)=p_{01}=|\textbf{Tr}(\rho |0\rangle\langle 1|)|. 
\end{equation}
Here, we provide a method to measure the quantity by measuring a overlap between two states. D. K. L. Oi ${\it et. al.}$ in \citep{oi1} gave the first proposal to measure various linear and nonlinear functions of density matrices in the interferometry directly. Later, the method was used in \cite{oi2} to measure various overlaps as shown in the Fig. (\ref{fig2}). 

\begin{figure}
\includegraphics[scale=0.15]{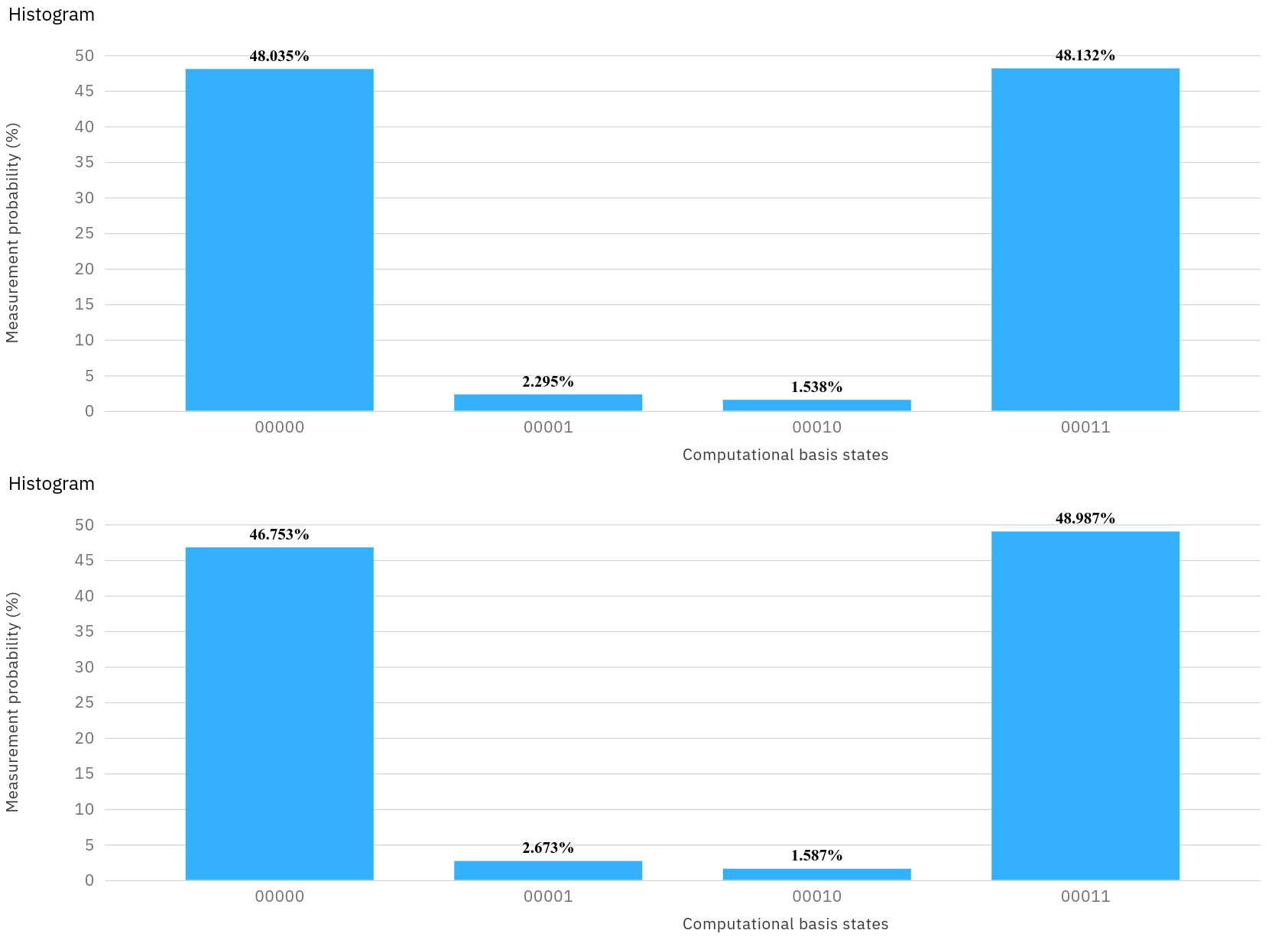}
\caption{
We implemented the circuit in Fig (\ref{fig2}) on the IBM quantum computer 5q-ibmq-vigo. It turns out that the entangled states generated by the IBM quantum computer maximally violates the inequality in Eq. (\ref{slhsineq}) or (\ref{equivalentof6}) for $i = j = \sigma_{z}$. Therefore, to measure the four terms of the Eq. (\ref{equivalentof5}) or (\ref{equivalentof6}), first we added the Pauli-x gate (Pauli-x is not available on the platform. Therefore, we implemented it by using the other two Pauli gates.) and then Pauli-y gate before both the detectors. We performed two runs of the experiment with 8192 shots each---one for each of the gates before the detectors. The probability of getting $|00\rangle$ (or $|11\rangle$) in the above figures is therefore, actually corresponds to the probability of getting the state $|++\rangle$ (or $|--\rangle$) when the Pauli-x is placed before the detectors. Thus, the entangled states produced by the IBM quantum computer violate the Eq. (\ref{slhsineq2}) and the left hand side reaches up to $~0.92$.
}
\label{fig3}
\end{figure}

We consider a state $\rho$ and measure its overlap with $|+\rangle=\frac{|0\rangle+|1\rangle}{\sqrt{2}}$ and $|i+\rangle=\frac{|0\rangle+i|1\rangle}{\sqrt{2}}$. We consider the value to be $p_{+}$ and $p_{i+}$ respectively. We can also measure the overlap with the state $|0\rangle$ and $|1\rangle$, which we consider to be $p_0$ and $p_1$. We define $a_{01}=\text{Tr}(\rho|0\rangle\langle 1|)$ and $a_{10}=\text{Tr}(\rho|1\rangle\langle 0|)$. One can easily show that $p_{+}=\frac{1}{2}(p_{0}+p_{1}+a_{01}+a_{10})$ and $p_{i+}=\frac{1}{2}(p_{0}+p_{1}-ia_{01}+ia_{10})$. Since, we know the values of $p_{+}$, $p_{i+}$, $p_{0}$ and $p_{1}$ from the measurements of overlaps with the corresponding states, we should be able to calculate the values of $a_{01}=a_{10}^{*}$ and $p_{01}$ by solving the two equations. 

However, one does not need to measure these local overlaps directly. Let us consider two states $|i^{a}\rangle$ and $|i^{a'}\rangle$ on the  Alice's side and similarly, two states $|j^{a}\rangle$ and $|j^{a'}\rangle$ $(a\neq a'$ and for two qubit scenario $a, a' \in \{1, 1\})$ on the Bob's side. Let us also form x-plus and x-minus states as $|i^{x\pm}\rangle = \frac{1}{\sqrt{2}}(|i^{a}\rangle\pm|i^{a'}\rangle)$. Similarly, y-plus and y-minus states are defined as $|i^{y\pm}\rangle = \frac{1}{\sqrt{2}}(|i^{a}\rangle\pm i |i^{a'}\rangle)$. One can easily show that the left hand sides of the inequalities in Eq. (\ref{slhsineq}) and \ref{slhsineq2} turn out to be
\begin{widetext}
\begin{eqnarray}
\underset{a\neq a'}{\sum_{a, a',}}p(|{i}^{a-a'}\rangle^{A},|{j}^{a-a'}\rangle^{B})&=&\langle i^{x+}j^{x+}|\rho|i^{x+}j^{x+}\rangle+\langle i^{x-}j^{x-}|\rho|i^{x-}j^{x-}\rangle-\langle i^{y+}j^{y+}|\rho|i^{y+}j^{y+}\rangle-\langle i^{y-}j^{y-}|\rho|i^{y-}j^{y-}\rangle
\label{equivalentof5}
\\ \text{and}\nonumber\\
\underset{a\neq a'}{\sum_{a, a',}}p(|{i}^{a-a'}\rangle^{A},|{j}^{a'-a}\rangle^{B})&=&\langle i^{x+}j^{x+}|\rho|i^{x+}j^{x+}\rangle+\langle i^{x-}j^{x-}|\rho|i^{x-}j^{x-}\rangle+\langle i^{y+}j^{y+}|\rho|i^{y+}j^{y+}\rangle\nonumber\\&+&\langle i^{y-}j^{y-}|\rho|i^{y-}j^{y-}\rangle-1.
\label{equivalentof6}
\end{eqnarray}
\end{widetext}
We use these identities for the two-qubit scenario to show the violation of the symmetric local hidden state model by the quantum mechanical states in the IBM quantum computer as depicted in the Fig (\ref{fig3}). However, showing the violation in the higher dimensions is a nontrivial task and at the same time, is of utmost importance for verifying our prediction.

\section{Self-testing}One of the ab initio motivations behind the wanderlust in the dark for an alternative description of $lhs$ model is of course to find a larger set of self-testable states and measurements, which can be used in various quantum information theoretic protocols. {\it A state $|\psi_{AB}\rangle$ and the measurement basis $|\phi^{a}\rangle$ $(a=0, 1, 2,....,$ a is the outcome of a measurement in the $\phi$  basis and ranges upto the dimension of the corresponding system$)$ are self-testable by the correlation measured via joint transition probability $p(|\phi^{a-a'}\rangle^{A}, |\phi^{b-b'}\rangle^{B})$ if all states $\rho_{AB}$ and measurements $|\widetilde{\phi}^{a}\rangle^{A}$ compatible with $p(|\phi^{a-a'}\rangle^{A}, |\phi^{b-b'}\rangle^{B})$ or $\langle\phi^{a}_{A}\phi^{b}_{B}|\rho_{AB}|\phi^{a'}_{A}\phi^{b'}_{B}\rangle$, turns out to be $|\psi_{AB}\rangle$ and the measurement $|\phi^{a}\rangle$}. 

To self-test one of the Bell states and measurement bases, we start with the assumptions that the transition amplitudes of an arbitrary state $\rho_{AB}$ in the measurement bases $|\phi^{a}\rangle^{A}$ and $|\xi^{a}\rangle^{A}$ take the maximal possible value of $\frac{1}{2}$, i.e.,
\begin{eqnarray}
\langle\phi^{0}_{A}\phi^{0}_{B}|\rho_{AB}|\phi^{1}_{A}\phi^{1}_{B}\rangle &=&\frac{1}{2}, \hspace{0.1cm}\langle\xi^{0}_{A}\xi^{0}_{B}|\rho_{AB}|\xi^{1}_{A}\xi^{1}_{B}\rangle=\frac{1}{2}\nonumber\\\langle\phi^{0}_{A}\phi^{1}_{B}|\rho_{AB}|\phi^{1}_{A}\phi^{0}_{B}\rangle &=&0,
\label{selftestassumption}
\end{eqnarray}
where 
$|\phi^{0}_{A(B)}\rangle$ and $|\phi^{1}_{A(B)}\rangle$ are two orthonormal states of Alice(Bob). 


We define $|\xi^{0}_{A(B)}\rangle=\frac{1}{\sqrt{2}}(|\phi^{0}_{A(B)}\rangle+ |\phi^{1}_{A(B)}\rangle)$ and $|\xi^{1}_{A(B)}\rangle=\frac{1}{\sqrt{2}}(|\phi^{0}_{A(B)}\rangle- |\phi^{1}_{A(B)}\rangle)$. We have the freedom to express the state $\rho_{AB}$ in the $|\phi^{i}_{A(B)}\rangle$ basis as $\rho_{AB} =\underset{i, k,\in S_{d_{A}}}{\sum_{j, l\in S_{d_B}}}\alpha_{ij}^{kl}|\phi^i_{A}\phi^j_{B}\rangle\langle \phi^k_{A}\phi^l_{B}|$ (without loss of generality), where $S_d=\{0, 1, 2, ....,d\}$. Here, $d$ is the dimension of the system. Hermiticity condition of the density matrix implies $\alpha^{kl}_{ij}=\alpha^{ij*}_{kl}$ and the normalization condition implies $\sum_{ij}\alpha^{ij}_{ij}=1$. First two equations and the Hermiticity condition imply that $\alpha_{11}^{00}=\alpha_{00}^{11}=\frac{1}{2}$ and $\alpha_{01}^{10}=\alpha_{10}^{01}=0$. The third equation together with the Hermiticity condition implies
$\underset{k, l}{\sum_{i, j=0}^{1}}(-1)^{k+l}\alpha_{ij}^{kl}=2$ and $\underset{k, l}{\sum_{i, j=0}^{1}}(-1)^{i+j}\alpha_{ij}^{kl}=2.$
Adding both the equations and putting the values of $\alpha_{11}^{00}=\alpha_{00}^{11}$ and $\alpha_{01}^{10}=\alpha_{10}^{01}$ we get,
\begin{equation}
\alpha^{00}_{00}-\alpha^{01}_{01}-\alpha^{10}_{10}+\alpha^{11}_{11}=1.
\label{wer}
\end{equation}
Subtracting the Eq. (\ref{wer}) from the normalization condition, we get
\begin{equation}
2\alpha^{01}_{01}+2\alpha^{10}_{10}+\sum_{i,j\neq 0,1}\alpha_{ij}^{ij} = 0.
\label{normaAndAddedEq}
\end{equation}
The fact that all the diagonal elements of a density matrix are non-negative and the Eq. (\ref{normaAndAddedEq}) imply $\alpha_{ij}^{ij}=0$ $\forall i, j$ except $\alpha_{00}^{00}$ and $\alpha_{11}^{11}$ and we get,
\begin{equation}
\alpha_{00}^{00}+\alpha_{11}^{11}=1.
\label{vitalpoint}
\end{equation}
Using the Cauchy-Schwartz inequality, one can show that 
\begin{equation}
\langle \phi^k\phi^l|\rho|\phi^i\phi^j\rangle|^2\leq |\langle \phi^i\phi^j|\rho|\phi^i\phi^j\rangle| |\langle \phi^k\phi^l|\rho|\phi^k\phi^l\rangle|.
\label{csineq}
\end{equation} 

To prove the inequality (\ref{csineq}), we start with
$|\langle \phi^k\phi^l|\rho|\phi^i\phi^j\rangle|^2=\langle \phi^k\phi^l|\rho|\phi^i\phi^j\rangle\langle \phi^i\phi^j|\rho|\phi^k\phi^l\rangle=\text{Tr}(|\phi^i\phi^j\rangle\langle \phi^i\phi^j|\rho|\phi^k\phi^l\rangle\langle \phi^k\phi^l|\rho)=\text{Tr}(A\rho B\rho)$, where we consider $A=|\phi^i\phi^j\rangle\langle \phi^i\phi^j|$ and $|\phi^k\phi^l\rangle\langle \phi^k\phi^l=B$. Now using the Cauchy-Schwarz inequality, one can show that $\text{Tr}(A\rho B\rho)\leq \sqrt{\text{Tr}(\rho A\rho A)\text{Tr}(\rho B\rho B)}$. This is nothing but the inequality (\ref{csineq}), which in turn implies, 
\begin{equation}
|\alpha^{kl}_{ij}|^{2}\leq |\alpha_{ij}^{ij}||\alpha_{kl}^{kl}|
\end{equation}

We use the inequality to show that for all vanishing diagonal elements, i.e., $\alpha_{ij}^{ij}=0$, all the off-diagonal elements are zero, i.e., $\alpha_{ij}^{kl}=\alpha_{kl}^{ij}=0 \forall i, j, k, l$.

Using the same inequality, one can also show that $\alpha_{00}^{00}\alpha_{11}^{11}\geq \frac{1}{4}$. From the bound together with the Eq. (\ref{vitalpoint}), one can show that 
\begin{equation}
(\alpha^{00}_{00}-\alpha^{11}_{11})^2\\=\alpha_{00}^{00^2}+\alpha_{11}^{11^2}
-2\alpha_{00}^{00}\alpha_{11}^{11}=1-4\alpha_{00}^{00}\alpha_{11}^{11}\leq 0.
\end{equation}

Since, $0\leq (\alpha_{00}^{00}-\alpha_{11}^{11})^2\leq 0$ implies $\alpha_{00}^{00}=\alpha_{11}^{11}$, we get $\alpha_{00}^{00}=\alpha_{11}^{11}=\frac{1}{2}$ from eq. (\ref{vitalpoint}). Using the fact that $\alpha_{01}^{01}=0$ in the above inequality, one can easily show that $|\alpha_{01}^{ij}|^2\leq 0$. This implies $\alpha_{01}^{ij}=\alpha^{01}_{ij}=0$ $\forall i, j.$  Therefore, we are left with terms $\alpha_{11}^{00}=\alpha_{00}^{11}=\alpha_{00}^{00}=\alpha_{11}^{11}=\frac{1}{2}$ implying that the state $\rho_{AB}$ is the Bell state $|\phi^{+}\rangle$ and the corresponding measurement bases are $z$ and $x$.

The next natural question to ask is possibly: Is it possible to extend the formalism of self-testing here for arbitrary states? A general proof for arbitrary two qubit entangled states of the form of $|\psi\rangle=\sqrt{\alpha}|00\rangle+\sqrt{1-\alpha}e^{i\phi}|11\rangle$ can indeed be extended from the simplified proof given here.
 
\section{Resource theory}We put forward a resource theory of this new form of nonlocality in this section. Just like other resources \cite{resource1,resource2,resource3, resource4} in quantum information theory, the first element we need to dig out is the free operations, i.e., operations under which a state belonging to $slhs$ model remains in $slhs$. Here we show that like entanglement or Bell nonlocality \cite{EPR, Schrodinger,Schrodinger1,bell}, this new form of nonlocality also cannot be created by local completely positive trace preserving operations $(lcptp)$ i.e.,
\begin{theorem}
a state $\rho_{ab}$ belonging to the $slhs$ model remains in the $slhs$ model under $lcptp$ operations $(\mathcal{LCPTP})$.
\end{theorem}
We lay down the proof of the theorem in the supplemental material \cite{supple}.

Once the free operations are found, the next logical step is to introduce an axiomatic approach to define the measures of the nonlocality. Finding the measure of a resource is a difficult task and even more difficult without knowing the conditions, it must satisfy. In this regard, we lay down a set of axioms, which must be satisfied by the measure $\mathcal{N}$ of the new form of nonlocality---$(a)$ $\mathcal{N}(\rho_{AB})=0$ $\forall \rho_{AB}\in slhs$, $(b) $ $\sum_{k}p_{k}\mathcal{N}(\mathcal{L}_{k}(\rho_{AB}))\leq \mathcal{N}(\rho_{AB})$ $\forall \mathcal{L}_{k}\in \mathcal{LCPTP}$, such that $\mathcal{L}_{k}(.)=\frac{r_{k}^{A}\otimes r_{k}^{B}(.)r_{k}^{A\dagger}\otimes r_{k}^{B\dagger}}{p_{k}}$, where $p_{k}=\textbf{Tr}(r_{k}^{A}\otimes r_{k}^{B}(.)r_{k}^{A\dagger}\otimes r_{k}^{B\dagger})$ and bipartite states $\rho_{AB}$. Additionally, it must not increase under classical mixing of bipartite states $\rho_{AB}^{i}$, i.e., $(c)$ $\mathcal{N}
(\rho_{AB})\leq\sum_{i}p_{i}\mathcal{N}(\rho_{AB}^{i})$, such that 
$\rho_{AB}=\sum_{i}p_{i}\rho_{AB}^{i}$ and $\sum_{i}p_{i}=1$.

In the supplemental material, we show that $|\langle ab|\rho_{AB}|cd\rangle|$ satisfies the axioms $(b)$ and $(c)$ for any bipartite state $\rho_{AB}$. By some numerical adjustments, we turn the quantity into a measure of the new form of non-locality.

\section{Conclision}  There was a formidable belief that the only way to describe the $lhs$ game is by giving up the freedom whether to trust or not to trust a party and retaining it for the other, thereby introducing an asymmetry in the game. In the $lhs$ game described earlier, Alice is considered to be an un-trusted party. Beauty of science is that it allows one to look at each and everything with suspicion, doubts and skepticism. Our skepticism about the game led to an alternative description of the $lhs$ model, which is symmetric, trust-less and can be used in self-testing the states and measurements. Therefore, it provides a better and operationally more efficient alternative to the Bell inequality for the self-testing and other device-independent protocols \cite{device,branciard,bennet}. 
 The symmetric form of $lhs$ model reveals more nonlocality than the existing $lhs$ or $lhv$ models (particularly in higher dimensions $d>2\otimes 2$) opening a new frontier for the experimentalists to test the new boundary of quantum correlation. In the two qubit scenario, an experimental violation of the inequalities can easily be shown and we performed an experiment in this scenario on the IBM quantum computer.


Bell nonlocality and $EPR$ nonlocality have already been verified experimentally \cite{tittel1,tittel2,nature,panexp1,zhang,zhao,weihs,optcomm,deng,cavalcanti,
steinlechner,hanchen,bennet,wittmann,smith,saunders,sun,armstrong,cmli} under different scenarios and assumptions in the last four decades. It will be really interesting to show the experimental violation of the inequalities given in this letter. Moreover, our study is limited to the bipartite scenario. Extending the formalism in the multipartite scenarios will be another interesting direction to explore.

{\it Acknowledgements---}DM and DK are supported by the National Research Foundation, Prime Minister’s Office, Singapore and the Ministry of Education, Singapore under the Research Centres of Excellence programme.

\newpage
\clearpage
\onecolumngrid
\section{Appendix}
\newcounter{defcounter}
\setcounter{defcounter}{0}
\newtheorem{defn}{Definition}
\newcommand{\comb}[2]{{}^{#1}\mathrm{C}_{#2}}
\newenvironment{myequationA}{%
\addtocounter{equation}{-1}
\refstepcounter{defcounter}
\renewcommand\theequation{A\thedefcounter}
\begin{equation}}
{\end{equation}}
\newenvironment{myequationB}{%
\addtocounter{equation}{-1}
\refstepcounter{defcounter}
\renewcommand\theequation{B\thedefcounter}
\begin{equation}}
{\end{equation}}
\newenvironment{myequationC}{%
\addtocounter{equation}{-1}
\refstepcounter{defcounter}
\renewcommand\theequation{C\thedefcounter}
\begin{equation}}
{\end{equation}}
\newenvironment{myequationD}{%
\addtocounter{equation}{-1}
\refstepcounter{defcounter}
\renewcommand\theequation{D\thedefcounter}
\begin{equation}}
{\end{equation}}
\newenvironment{myequationE}{%
\addtocounter{equation}{-1}
\refstepcounter{defcounter}
\renewcommand\theequation{E\thedefcounter}
\begin{equation}}
{\end{equation}}
\newenvironment{myequationF}{%
\addtocounter{equation}{-1}
\refstepcounter{defcounter}
\renewcommand\theequation{F\thedefcounter}
\begin{equation}}
{\end{equation}}
\newenvironment{myequationG}{%
\addtocounter{equation}{-1}
\refstepcounter{defcounter}
\renewcommand\theequation{G\thedefcounter}
\begin{equation}}
{\end{equation}}
\newenvironment{myequationH}{%
\addtocounter{equation}{-1}
\refstepcounter{defcounter}
\renewcommand\theequation{H\thedefcounter}
\begin{equation}}
{\end{equation}}
\newenvironment{myeqnarrayA}{%
\addtocounter{equation}{-1}
\refstepcounter{defcounter}
\renewcommand\theequation{A\thedefcounter}
\begin{eqnarray}}
{\end{eqnarray}}
\newenvironment{myeqnarrayB}{%
\addtocounter{equation}{-1}
\refstepcounter{defcounter}
\renewcommand\theequation{B\thedefcounter}
\begin{eqnarray}}
{\end{eqnarray}}
\newenvironment{myeqnarrayC}{%
\addtocounter{equation}{-1}
\refstepcounter{defcounter}
\renewcommand\theequation{C\thedefcounter}
\begin{eqnarray}}
{\end{eqnarray}}
\newenvironment{myeqnarrayD}{%
\addtocounter{equation}{-1}
\refstepcounter{defcounter}
\renewcommand\theequation{D\thedefcounter}
\begin{eqnarray}}
{\end{eqnarray}}
\newenvironment{myeqnarrayE}{%
\addtocounter{equation}{-1}
\refstepcounter{defcounter}
\renewcommand\theequation{E\thedefcounter}
\begin{eqnarray}}
{\end{eqnarray}}
\newenvironment{myeqnarrayF}{%
\addtocounter{equation}{-1}
\refstepcounter{defcounter}
\renewcommand\theequation{F\thedefcounter}
\begin{eqnarray}}
{\end{eqnarray}}
\newenvironment{myeqnarrayG}{%
\addtocounter{equation}{-1}
\refstepcounter{defcounter}
\renewcommand\theequation{G\thedefcounter}
\begin{eqnarray}}
{\end{eqnarray}}
\newenvironment{myeqnarrayH}{%
\addtocounter{equation}{-1}
\refstepcounter{defcounter}
\renewcommand\theequation{H\thedefcounter}
\begin{eqnarray}}
{\end{eqnarray}}

\section*{Stronger SLHS model} A stronger $slhs$ model would be just to express the complex off-diagonal terms corresponding to the transition probabilities of a bipartite state in terms of the complex off-diagonal terms of hidden states,
\begin{myequationA}
\langle i^{a}i^{b}|\rho_{AB}|i^{a'}i^{b'}\rangle=\sum_{\lambda}p(\lambda)\langle i^{a}|\rho^{\lambda}_{A}|i^{a'}\rangle\langle i^{b}|\rho^{\lambda}_{B}|i^{b'}\rangle,
\label{strongerslhs}
\end{myequationA}
where $\rho^{\lambda}_{A}$ and $\rho^{\lambda}_{B}$ are hidden states of Alice and Bob respectively with a distribution $p(\lambda)$ prepared by the unknown source to cheat them. 

The inequality in Eq. (\ref{strongerslhs}) is stronger than the Eq. (4) of the article is due to the fact that the former implies the later but not the other way around. To prove it, we start with,
\begin{eqnarray}
|\langle i^{a}i^{b}|\rho_{AB}|i^{a'}i^{b'}\rangle|&\leq &\sum_{\lambda}p(\lambda)|\langle i^{a}|\rho^{\lambda}_{A}|i^{a'}\rangle||\langle i^{b}|\rho^{\lambda}_{B}|i^{b'}\rangle|,\nonumber
\end{eqnarray}
where we use the Eq. (\ref{strongerslhs}) and the triangle inequality. 
The inequality above implies that there must exist a set of hidden states $\sigma_{A}^{\lambda}$ and $\sigma_{B}^{\lambda}$ with a distribution $p'(\lambda)$ such that
\begin{eqnarray}
|\langle i^{a}i^{b}|\rho_{AB}|i^{a'}i^{b'}\rangle|=\sum_{\lambda}p'(\lambda)|\langle i^{a}|\sigma^{\lambda}_{A}|i^{a'}\rangle||\langle i^{b}|\sigma^{\lambda}_{B}|i^{b'}\rangle|,\nonumber
\end{eqnarray}
which is nothing but the Eq. (4) of the letter.

\section*{Proof of Eq. 8.}
In the {\bf symmetric local hidden state model} section of the main article, we have given two inequalities in Eq. (8) and (9) for general bipartite scenario. The inequalities do not follow directly from the proof given in Eq. (5) for the two-qubit scenario. Here we provide the general proof. We start with the joint transition probability as 
\begin{myeqnarrayA}
&p&(|i^{a-a'}\rangle^{A},|j^{a-a'}\rangle^{B})=\sum_{\lambda}p(\lambda)p\bigg{(}|{i}^{a-a'}\rangle^{A}|\lambda\bigg{)}p\bigg{(}|{j}^{a-a'}\rangle^{B}|\lambda\bigg{)}\nonumber\\ &\leq &   \sqrt{\sum_{\lambda'}p(\lambda')p^{2}\bigg{(}|i^{a-a'}\rangle^{A}\bigg{|}\lambda'\bigg{)}\sum_{\lambda}p(\lambda)p^{2}\bigg{(}|j^{a-a'}\rangle^{B}\bigg{|}\lambda\bigg{)}}\nonumber\\& =& \sqrt{\overline{p^{2}\bigg{(}|i^{a-a'}\rangle^{A}\bigg{)}}. \overline{p^{2}\bigg{(}|j^{a-a'}\rangle^{B}\bigg{)}}}.
\end{myeqnarrayA} 
Here, we used the fact that the transition probability $p(|i^{a-a'}\rangle)$ is bounded from above by $\frac{1}{2}$ in the derivation of the two-qubit inequality in Eq. (5). In the higher dimensions, where $a$ and $a'$ have more than two values $0$ and $1$, $p(|i^{0-1}\rangle)=\frac{1}{2}$ implies $p(|i^{1-2}\rangle)=p(|i^{0-2}\rangle)=0$. Therefore, for $p(|i^{a-a'}\rangle=\frac{1}{2})$, the maximal value cannot be reached for local dimensions more than two. The fact that $\sum_{t}\max_{x}f(t,x)$ will not necessarily be greater than $\max_{x}\sum_{t}f(t,x)$ for any $f(t,x)$ and we are supposed to find the maximum value of a quantity of the form of $\max_{x}\sum_{t}f(t,x)$, implies we need to take an alternative route. We need to calculate the maximum of $\sum_{a,a'}\sqrt{\overline{p^{2}\bigg{(}|i^{a-a'}\rangle^{A}\bigg{)}}. \overline{p^{2}\bigg{(}|j^{a-a'}\rangle^{B}\bigg{)}}}$ for arbitrary states $\rho_{A}$ and $\sigma_B$. One can show that the maxima can be found when $p(|i^{a-a'}\rangle^{A})=\frac{1}{d}  \text{$\forall a$ and $a'$}$. Therefore,
\begin{myeqnarrayA}
\max \sum_{a, a'}\sqrt{\overline{p^{2}\bigg{(}|i^{a-a'}\rangle^{A}\bigg{)}}. \overline{p^{2}\bigg{(}|j^{a-a'}\rangle^{B}\bigg{)}}}=\frac{d-1}{d}.
\end{myeqnarrayA} 
This proves the inequality in Eq. (8) in the article.
\section*{Proof of Theorem 1.}
We consider a state $\rho_{AB}$ belongs to the set of $slhs$ states, such that 
\begin{widetext}
\begin{myequationA}
p\bigg{(}|i^{a-a'}\rangle,|j^{a-a'}\rangle\bigg{)}=|\langle i^aj^a|\rho_{AB}|i^{a'}j^{a'}\rangle|\overset{SLHS}{=}\sum_{\lambda}p(\lambda)|\langle i^{a}|\rho_{A}^{\lambda}|i^{a'}\rangle||\langle j^{a}|\rho_{B}^{\lambda}|j^{a'}\rangle|,
\label{freeopcond}
\end{myequationA}
\end{widetext}
for any basis labeled by $i$ and a particular vector from the basis labeled by $a$, such that $a\neq a'$. The state $\rho_{AB}$, under local $cptp$ operation, transforms to the state $\rho'_{AB}$, such that $\rho'_{AB}=\mathcal{L}(\rho_{AB})=\sum_{k}r_{k}^{A}\otimes r_{k}^{B}\rho_{AB}r_{k}^{A\dagger}\otimes r_{k}^{B\dagger}$. We define $\mathcal{L}_{k}(\rho_{AB})=\frac{r_{k}^{A}\otimes r_{k}^{B}\rho_{AB}r_{k}^{A\dagger}\otimes r_{k}^{B\dagger}}{\text{Tr}(r_{k}^{A}\otimes r_{k}^{B}\rho_{AB}r_{k}^{A\dagger}\otimes r_{k}^{B\dagger})}$, $p_{k}=\text{Tr}(r_{k}^{A}\otimes r_{k}^{B}\rho_{AB}r_{k}^{A\dagger}\otimes r_{k}^{B\dagger})$, such that $\sum_{k}p_{k}=1$ and $\sum_{k}r_{k}^{\dagger}r_{k}\leq I$ for both the systems. Here, we show that the state $\rho'_{AB}$ also belongs to the set of $SLHS$ states. We start with the quantity $\langle i^a i^a|\rho'_{AB}|i^{a'}i^{a'}\rangle$ such that $a\neq a'$
\begin{widetext}
\begin{myeqnarrayA}
\langle i^{a} j^{a}|\rho'_{AB}|i^{a'} j^{a'}\rangle&=&\sum_{k}\langle i^a j^a|r_{k}^{A}\otimes r_{k}^{B}\rho_{AB}r_{k}^{A\dagger}\otimes r_{k}^{B\dagger}|i^{a'} j^{a'}\rangle\nonumber = \sum_{k}\text{Tr}\Big{(}r_{k}^{A}\otimes r_{k}^{B}\rho_{AB}r_{k}^{A\dagger}\otimes r_{k}^{B\dagger}|i^{a'}j^{a'}\rangle\langle i^aj^a|\Big{)}\nonumber\\&=&\sum_{k}\text{Tr}\Big{(}\rho_{AB}A_{k}\otimes B_{k}\Big{)}=\sum_{k,m,n,p,q}\langle mn|\rho_{AB}|pq\rangle \langle pq|A_{k}\otimes B_{k}|mn\rangle,\nonumber\\ && \vspace{1cm}\text{where $A_{k}=r_{k}^{A\dagger}| i^{a'}\rangle\langle i^{a}|r_{k}^{A}$ and $B_k=r_{k}^{B\dagger}| j^{a'}\rangle\langle j^{a}|r_{k}^{B}$}\nonumber\\&=&\sum_{k,m,n,p,q,\lambda}\langle p|A_k|m\rangle\langle q|B_{k}|n\rangle p(\lambda)\langle m |\rho_{A}^{\lambda}|p\rangle\langle n|\rho_{B}^{\lambda}|q\rangle=\sum_{k,p,q,\lambda}p(\lambda)\langle p |A_{k}\rho^{\lambda}_{A}|p\rangle\langle q|B_{k}\rho_{B}^{\lambda}|q\rangle\nonumber\\&=&\sum_{k,\lambda}
p(\lambda)\langle i^a|r_{k}^{A}\rho_{A}^{\lambda}r_{k}^{A\dagger}|i^{a'}\rangle\langle j^a|r_{k}^{A}\rho_{B}^{\lambda}r_{k}^{B\dagger}|j^{a'}\rangle.
\label{freeop}
\end{myeqnarrayA}
This proof is for the stronger version of our $slhs$ model. To prove it for the weaker version, we proceed further with the following,
\begin{myeqnarrayA}
\Big{|}\langle i^{a} j^{a}|\rho'_{AB}|i^{a'} j^{a'}\rangle \Big{|}&=&\Big{|}\sum_{k,\lambda}
p(\lambda)\langle i^a|r_{k}^{A}\rho_{A}^{\lambda}r_{k}^{A\dagger}|i^{a'}\rangle\langle j^a|r_{k}^{A}\rho_{B}^{\lambda}r_{k}^{B\dagger}|j^{a'}\rangle\Big{|}\leq \sum_{k,\lambda}
p(\lambda)\Big{|}\langle i^a|r_{k}^{A}\rho_{A}^{\lambda}r_{k}^{A\dagger}|i^{a'}\rangle\langle j^a|r_{k}^{A}\rho_{B}^{\lambda}r_{k}^{B\dagger}|j^{a'}\rangle \Big{|}\nonumber\\&=&\sum_{k,\lambda}
p(\lambda)\Big{|}\langle i^a|r_{k}^{A}\rho_{A}^{\lambda}r_{k}^{A\dagger}|i^{a'}\rangle\Big{|}\Big{|}\langle j^a|r_{k}^{A}\rho_{B}^{\lambda}r_{k}^{B\dagger}|j^{a'}\rangle \Big{|}
\label{freeop1}
\end{myeqnarrayA}
\end{widetext}
which implies that the state $\rho_{AB}'$ also belongs to the set of $slhs$. In the equation above in Eq. (\ref{freeop}), in the second line, we use the fact that $r_{k}^{A(B)}|i^{a'}\rangle \langle i^{a}|r_{k}^{A(B)\dagger}=A_{k}(B_{k})$, in the third line, we use the fact that the state $\rho_{AB}\in slhs$, i.e., the Eq. (\ref{strongerslhs}). In the first inequality in Eq. (\ref{freeop1}), we use the triangle inequality for complex numbers.

\section{Proof of Axioms.}
We consider  $\mathcal{N}(\rho_{AB})=\underset{a \neq a', b\neq b'}{\sum_{a, a', b, b'}}\max\Bigg{\{}\max_{i, j}|\langle i^aj^b|\sum_{i}p_{i}\rho_{AB}^{i}|i^{a'}j^{b'}\rangle|-\frac{1}{d^2},0\Bigg{\}}$ to be a measure of the new form of $EPR$ correlation, where $d$ is the dimension of the subsystems.
To prove the axiom (b), we start with the fact that 
\begin{myeqnarrayA}
|\langle a b|\mathcal{L}_{k}^{AB}\rho_{AB}|c d\rangle|=\frac{|\sum_{n,m,p,q}\langle mn |\rho_{AB}|pq\rangle\langle pq | A_{k}\otimes B_{k}|mn\rangle\nonumber|}{p_{k}},
\end{myeqnarrayA}

where $A_{k}=r^{A\dagger}_{k}|c\rangle\langle a|r_{k}^{A}$ and $B_{k}=r^{B\dagger}_{k}|d\rangle\langle b|r_{k}^{B}$. Therefore,
\begin{widetext}
\begin{myeqnarrayA}
\sum_{k}p_{k}|\langle a b|\mathcal{L}_{k}^{AB}(\rho_{AB})|c d\rangle|&\leq & \sum_{n,m,p,q,k}|\langle mn |\rho_{AB}|pq\rangle\langle pq | A_{k}\otimes B_{k}|mn\rangle|\nonumber\\&=&\sum_{n,m,p,q,k}|\langle mn |\rho_{AB}|pq\rangle||\langle p | A_{k}|m\rangle||\langle q| B_{k}|n\rangle|,
\label{proof1-1}
\end{myeqnarrayA}
\end{widetext}
where in the inequality, we use the triangle inequality. Now, we start with the quantity $|\langle p | A_{k}|m\rangle|$ as
\begin{myeqnarrayA}
&&\sum_{k}|\langle p | A_{k}|m\rangle|=\sum_{k}|\langle p|r^{A\dagger}_{k}|c\rangle\langle a|r_{k}^{A}|m\rangle|\nonumber\\ &&=\sum_{k}|[r^{A\dagger}_{k}]_{cp}[r^{A}_{k}]_{ma}|\nonumber\\&&=\sum_{k}|[r^{A\dagger}_{k}]_{cp}[r^{A}_{k}]_{pa}|+\underset{p\neq m}{\sum_{k}}|[r^{A\dagger}_{k}]_{cp}[r^{A}_{k}]_{ma}|,
\label{proof1-2}
\end{myeqnarrayA}
where the first term is zero due to the fact that $\sum_{k}r^{\dagger}_{k}r_{k}=I_{d}$ and off-diagonal terms of an identity matrix are all zero. We start with the second term as
\begin{myeqnarrayA}
&&\underset{p\neq m}{\sum_{k}}|[r^{A\dagger}_{k}]_{cp}[r^{A}_{k}]_{ma}|\leq\sum_{k}|[r^{A\dagger}_{k}]_{cp}[r^{A}_{k}]_{ma}|\nonumber\\ &\leq &\sqrt{\sum_{k}|[r_{k}^{A\dagger}]_{cp}|^2\sum_{l}|[r_{l}^{A}]_{ma}|^2}.
\label{proof1-3}
\end{myeqnarrayA}
Now, we start with the first term under the square-root, i.e.,
\begin{myeqnarrayA}
\sum_{k}|[r_{k}^{A\dagger}]_{cp}|^2&=&\sum_{k}\langle c|r^{A\dagger}_{k}|p\rangle\langle p|r_{k}^{A}|c\rangle.
\label{proof1-4}
\end{myeqnarrayA}
We know $\text{Tr}(\sum_{k}r_{k}^{A}\rho_{0}r_{k}^{A\dagger})=1$ and $\text{Tr}(r_{k}^{A}\rho_{0}r_{k}^{A\dagger})=p_{k}$ such that $\sum_{k}p_{k}=1$ for any 
arbitrary state $\rho_{0}$. We consider $\rho_{0}=|c\rangle\langle c|$. Therefore, we get
\begin{myeqnarrayA}
p_{k}&=&\text{Tr}(r_{k}^{A}\rho_{0}r_{k}^{A\dagger})=\sum_{m}\langle m|r_{k}^{A}|c\rangle\langle c| r_{k}^{A\dagger}|m\rangle\nonumber \\\text{or,}&&\langle m|r_{k}^{A}|c\rangle\langle c|r_{k}^{A\dagger}|m\rangle \leq p_{k},
\label{proof1-5}
\end{myeqnarrayA} 
where $|m\rangle$ is an arbitrary state and forms a complete basis and the inequality comes from the fact that each term within the summation is positive.  Now, we start with the same quantity again to show that
\begin{myeqnarrayA}
p_{k}&=&\sum_{m}\langle m|r_{k}^{A}|c\rangle\langle c| r_{k}^{A\dagger}|m\rangle\nonumber\\&=&\sum_{m}\langle m|r_{k}^{A\dagger}r_{k}^{A}|c\rangle\langle c| |m\rangle=\langle c|r_{k}^{A\dagger}r_{k}^{A}|c\rangle.
\label{proof1-6}
\end{myeqnarrayA}
Thus, from Eq. (\ref{proof1-5}) and (\ref{proof1-6}), we get
\begin{myeqnarrayA}
|\langle m|r_{k}^{A}|c\rangle|^2\leq p_{k}=\delta_{mc}\langle m|r_{k}^{A\dagger}r_{k}^{A}|c\rangle
\label{proof1-7}
\end{myeqnarrayA}
for any states $|m\rangle$ and $|c\rangle$. Now, from Eq. (\ref{proof1-7}) and (\ref{proof1-4}), we get
 \begin{myeqnarrayA}
 \sum_{k}|[r_{k}^{A\dagger}]_{cp}|^2\leq \delta_{cp}.
 \label{proof1-8}
 \end{myeqnarrayA}
Using Eq. (\ref{proof1-1}), (\ref{proof1-2}), (\ref{proof1-3}) and (\ref{proof1-8}), we get

\begin{myeqnarrayA}
\sum_{k}p_{k}|\langle i^a i^b|\frac{\mathcal{L}_{k}^{AB}(\rho_{AB})}{p_{k}}|i^c i^d\rangle|\leq |\langle i^a i^b|\rho_{AB}|i^c i^d\rangle|,
\end{myeqnarrayA}
which proves the axiom (b).

Now, we turn to the axiom (c). We consider a bipartite state $\rho_{AB}$ such that $\rho_{AB}=\sum_{i}p_{i}\rho_{AB}^{i}$, where $\sum_{i}p_{i}=1$. We start with the following
\begin{myeqnarrayA}
&|\langle ab|\sum_{i}p_{i}\rho_{AB}^{i}|cd\rangle|=|\sum_{i}p_{i}\langle ab|\rho_{AB}^{i}|cd\rangle|\nonumber\\&\leq \sum_{i}|p_{i}\langle ab |\rho_{AB}^{i}|cd\rangle|=\sum_{i}p_{i}|\langle ab|\rho_{i}|cd\rangle|.
\end{myeqnarrayA}
Thus, we prove that $\mathcal{N}(\rho_{AB})\leq\sum_{i}p_{i}\mathcal{N}(\rho_{AB}^{i})$.
\section*{No separable state can violate the inequalities}
We consider the transition probability of a bipartite, two-qubit state from $|0\rangle$ to $|1\rangle$ on Alice and as well as Bob's side as $p(|{z}^{0-1}\rangle^{A},|{z}^{0-1}\rangle^{B})$, where $|z^0\rangle$ or $|z^1\rangle$ stands for the states $|0\rangle$ or $|1\rangle$ respectively. For a two-qubit separable state $\rho_{AB}$, the transition probability is given by
\begin{myeqnarrayA}
&p(|{z}^{0-1}\rangle^{A},|{z}^{0-1}\rangle^{B})=\langle 00|\rho_{AB}|11\rangle\langle 11|\rho_{AB}|00\rangle\nonumber\\&=\sum_{\lambda, \lambda'}p_{\lambda}p_{\lambda'}\rho^{A}_{01}(\lambda)\rho^{B}_{01}(\lambda)\rho^{A}_{10}(\lambda')\rho^{B}_{10}(\lambda')\nonumber\\&\leq\sum_{\lambda}p_{\lambda}|\rho^{A}_{01}(\lambda)|^2\sum_{\lambda'}p_{\lambda'}|\rho^{B}_{01}(\lambda')|\leq\frac{1}{4}.
\end{myeqnarrayA}
Therefore, two terms $p(|{z}^{0-1}\rangle^{A},|{z}^{0-1}\rangle^{B})$ and $p(|{z}^{1-0}\rangle^{A},|{z}^{1-0}\rangle^{B})$ can at most add upto $\frac{1}{2}$ and cannot go beyond that value. The arguments here can be extended even for the general bipartite separable states. Therefore, separable states cannot violate the inequality given in Eq. (5)-(9) in the main article. Does the same hold for the existing one-sided LHS model as well? We show that the answer to this question to be affirmative. 

To prove it, we start with the steering game, where Alice prepares a two-qubit bipartite state. She sends one part of it to Bob and keeps another part with her. She claims that his state is entangled with her and she can steer his state. Bob does not believe Alice. He asks Alice to perform measurements in certain bases. After Alice's measurements in each basis, Bob measures certain quantities on his system and he claims that the (conditional) states can in principle be explained by hidden state model as laid down in Eq. (1). The overall state of Alice and Bob can be written as (Bob's assumption)
\begin{myequationA}
\rho_{AB}=\sum_{a, \theta}p(a,\theta)|a\rangle\langle a|\otimes\rho_{a|\Pi(\theta)}.
\end{myequationA}
Now on this state, if Alice and Bob really performs the measurements as prescribed in the article, one can easily show following the previous arguments for separable states that such states cannot violate the inequality based on the  symmetric hidden state model.


\begin{thebibliography}{99}
\bibitem{bell}J. S. Bell, {\it On the Einstein Podolsky Rosen paradox}, Physics {\bf 1}, 195 (1964); 
\bibitem{jfclauser}J. F. Clauser, M. A.
Horne, A. Shimony, and R. A. Holt, {\it Proposed Experiment to Test Local Hidden-Variable Theories},
 Phys. Rev. Lett. {\bf 23},
880 (1969).
\bibitem{lhsintro} H. M. Wiseman, S. J. Jones, and A. C. Doherty, {\it Steering, Entanglement, Nonlocality, and the Einstein-Podolsky-Rosen Paradox}, Phys.
Rev. Lett. 98, 140402 (2007).
\bibitem{sjjones} S. J. Jones, H. M. Wiseman,
and A. C. Doherty, Entanglement, {\it Einstein-Podolsky-Rosen correlations, Bell nonlocality, and steering}, Phys. Rev. A {\bf 76}, 052116 (2007).
\bibitem{wittmann} B. Wittmann, S. Ramelow, F. Steinlechner, N. K. Lang?ford, N. Brunner, H. M. Wiseman, R. Ursin, and
A. Zeilinger, {\it Loophole-free Einstein–Podolsky–Rosen experiment
via quantum steering}, New J. of Phys. {\bf 14}, 053030 (2012).

\bibitem{nguyen}H. C. Nguyen, H.-V. Nguyen, and O. G\"uhne, {\it Geometry of Einstein-Podolsky-Rosen Correlations}, Phys. Rev. Lett. {\bf 122}, 240401 (2019).
\bibitem{uola}R. Uola, A. C. S. Costa, H. C. Nguyen,
and O. G\"uhne, {\it Quantum Steering}, Rev. Mod. Phys. {\bf 92}, 15001 (2020).


\bibitem{resource2}R. Gallego, L. Aolita, {\it Resource Theory of Steering}, Phys. Rev. X {\bf 5}, 041008 (2015).
\bibitem{resource3}A. Streltsov, G. Adesso, M. B. Plenio, {\it Colloquium: Quantum coherence as a resource}, Rev. Mod. Phys. {\bf 89}, 041003 (2017).

\bibitem{jones}S. J. Jones, H. M. Wiseman, A. C. Doherty, {\it Entanglement, Einstein-Podolsky-Rosen correlations, Bell nonlocality, and steering}, Phys. Rev. A {\bf 76}, 052116 (2007).
\bibitem{pati}J.-L. Chen, H.-Y. Su, Z.-P. Xu, A. K. Pati, {\it Sharp Contradiction for Local-Hidden-State Model in Quantum Steering}, Sci. Rep. {\bf 6}, 32075 (2016).
\bibitem{device}A. A\'cin, N. Brunner, N. Gisin, S. Massar, S. Pironio, and
V. Scarani, {\it Device-Independent Security of Quantum Cryptography against Collective Attacks}, Phys. Rev. Lett. {\bf 98}, 230501 (2007).
\bibitem{branciard}C. Branciard,  E. G. Cavalcanti, S. P. Walborn, V. Scarani,
and H. M. Wiseman, {\it One-sided device-independent quantum key distribution: Security, feasibility, and the connection with steering
}, Phys. Rev. A {\bf 85}, 010301(R) (2012).
\bibitem{bennet}Bennett C. H., G. Brassard, and N. D. Mermin, {\it Quantum cryptography without Bell’s theorem}, Phys.
Rev. Lett. {\bf 68}, 557(1992).
\bibitem{supple} The Supplemental material
\bibitem{deba1} D. Mondal, T. Pramanik, A. K. Pati, {\it Nonlocal advantage of quantum coherence}, Phys. Rev. A 95,
010301(R) (2017).
\bibitem{deba2} D. Mondal, D. Kaszlikowski, {\it Complementarity Relations Between Quantum Steering Criteria}, Phys. Rev. A {\bf 98},
052330(2018).
\bibitem{deba3}D. Mondal, C. Datta, J. Singh, D. Kaszlikowski, {\it Authentication protocol based on collective quantum steering}, Phys. Rev. A {\bf 99}, 012312 (2019).
\bibitem{deba4}D. Mondal, J. Singh, D. Kaszlikowski, {\it No nonlocal advantage of quantum coherence beyond quantum instrumentality}, arXiv:1901.07008.
\bibitem{saunders} D. J. Saunders, S. J. Jones, H. M. Wiseman, and G. J. Pryde, {\it Experimental EPR-Steering of Bell-local States}, 
Nat. Phys. {\bf 6}, 845 (2010).
\bibitem{eberle} V. Händchen, T. Eberle, S. Steinlechner, A. Samblowski, T.
Franz, R. F. Werner, and R. Schnabel, {\it Observation of one-way Einstein-Podolsky-Rosen steering
}, Nat. Photonics {\bf 6}, 596
(2012).
\bibitem{smith} D. H. Smith, G. Gillett, M. de Almeida, C. Branciard,
A. Fedrizzi, T. J. Weinhold, A. Lita, B. Calkins, T. Gerrits,
H. M. Wiseman, S. W. Nam, and A. G. White, {\it Conclusive quantum steering with superconducting transition-edge sensors}, Nat. Commun. {\bf 3}, 625 (2012).
\bibitem{bennet1} A. J. Bennet, D. A. Evans, D. J. Saunders, C. Branciard,
E. G. Cavalcanti, H. M. Wiseman, and G. J. Pryde, {\it Arbitrarily Loss-Tolerant Einstein-Podolsky-Rosen Steering Allowing a Demonstration over 1 km of Optical Fiber with No Detection Loophole}, Phys.
Rev. X {\bf 2}, 031003 (2012).
\bibitem{tischler} N. Tischler, F. Ghafari, T. J. Baker, S. Slussarenko, R. B.
Patel, M. M. Weston, S. Wollmann, L. K. Shalm, V. B.
Verma, S. W. Nam, H. C. Nguyen, H. M. Wiseman, and
G. J. Pryde, {\it Conclusive Experimental Demonstration of One-Way Einstein-Podolsky-Rosen Steering}, Phys. Rev. Lett. {\bf 121}, 100401 (2018).
 \bibitem{ksun}K. Sun, X.-J. Ye, J.-S. Xu, X.-Y. Xu, J.-S. Tang, Y.-C. Wu,
J.-L. Chen, C.-F. Li, and G.-C. Guo, {\it Experimental Quantification of Asymmetric Einstein-Podolsky-Rosen Steering
}, Phys. Rev. Lett. {\bf 116},
160404 (2016).
\bibitem{yxiao}Y. Xiao, X.-J. Ye, K. Sun, J.-S. Xu, C.-F. Li, and G.-C. Guo,
 {\it Demonstration of Multisetting One-Way Einstein-Podolsky-Rosen Steering in Two-Qubit Systems}, Phys. Rev. Lett. 118, 140404 (2017).
\bibitem{oi1}A. K. Ekert, C. M. Alves, D. K. L. Oi, M. Horodecki, P.
Horodecki, and L. C. Kwek, {\it Direct Estimations of Linear and Nonlinear Functionals of a Quantum State}, Phys. Rev. Lett. {\bf 88}, 217901 (2002).
\bibitem{oi2} K. Bartkiewicz, K. Lemr and A. Miranowicz, {\it Direct method for measuring of purity, superfidelity, and subfidelity of photonic two-qubit mixed states
}, Phys. Rev. A {\bf 88},
052104 (2013).
\bibitem{resource1}E. Wolfe, D. Schmid, A. B. Sainz, R. Kunjwal, R. W. Spekkens, {\it Quantifying Bell: the Resource Theory of Nonclassicality of Common-Cause Boxes}, Quantum {\bf 4}, 280 (2020).
\bibitem{resource4}J. H. Selby and C. M. Lee, {\it Compositional resource theories of coherence}, Quantum {\bf 4}, 319 (2020).

\bibitem{Schrodinger}E. Schrödinger, {\it Discussion of Probability Relations between
Separated Systems}, Proc. Cambridge Philos. Soc. {\bf 31}, 553-563
(1935); 
\bibitem{Schrodinger1} E. Schrödinger, {\it Probability relations between separated systems}, Proc. Cambridge Philos. Soc. {\bf 32},
446 (1936).
\bibitem{EPR}A. Einstein, D. Podolsky, N. Rosen, {\it Can Quantum-Mechanical Description of Physical Reality Be Considered Complete?}, Phys. Rev. {\bf 47}, 777
(1935).
\bibitem{aspect}A. Aspect, J. Dalibard, and G. Roger, {\it Experimental Test of Bell's Inequalities Using Time-Varying Analyzers}, Phys. Rev. Lett. {\bf 49}, 1804 (1982).
\bibitem{tittel1} W. Tittel, J. Brendel, H. Zbinden, and N. Gisin, {\it Violation of Bell Inequalities by Photons More Than 10 km Apart},
Phys. Rev. Lett. 81, 3563 (1998).
\bibitem{tittel2}W. Tittel, J. Brendel, B. Gisin, T. Herzog, H. Zbinden, and N. Gisin, {\it Experimental demonstration of quantum correlations over more than 10 km},
Phys. Rev. A {\bf 57}, 3229 (1998).
\bibitem{weihs}G. Weihs, T. Jennewein, C. Simon, H. Weinfurter, and A. Zeilinger, {\it Violation of Bell's Inequality under Strict Einstein Locality Conditions},
Phys. Rev. Lett. {\bf 81}, 5039 (1998).
\bibitem{zhang}C. Zhang, C.-J. Zhang, Y.-F. Huang, Z.-B. Hou, B.-H. Liu, C.-F. Li and G.-C. Guo, {\it Experimental test of genuine multipartite nonlocality under the no-signalling principle
}, Sci. Reps. {\bf 6}, 39327(2016).
\bibitem{panexp1}Z. Zhao, T. Yang, Y.-A. Chen, A.-N. Zhang, M. Zukowski, and  J.-W. Pan, {\it Experimental Violation of Local Realism by Four-Photon Greenberger-Horne-Zeilinger Entanglement}, Phys. Rev. Lett. {\bf 91}, 180401 (2003).
\bibitem{zhao}J.-Q. Zhao, L.-Z. Cao, Y. Yang, Y.-D. Li, H.-X. Lu, {\it Experimental verification of a new Bell-type inequality}, Phys. Lett. A {\bf 382}, 1214 (2018).
\bibitem{nature}C. Abellán, A. Acín, A. Alarcón, O. Alibart, C. K. Andersen, F. Andreoli, A. Beckert, F. A. Beduini, A. Bendersky, et. al., {\it Challenging local realism with human choices}, Nat. {\bf 557}, 212(2018).
\bibitem{optcomm}C.-M. Li, H.-P. Lo,  L.-Y. Chen, A. Yabushita, {\it Experimental verification of multidimensional quantum steering}, Opt. Comm. {\bf 410}, 956 (2018).


\bibitem{saunders} D. Saunders, S. Jones, H. Wiseman, and G. Pryde, {\it Erratum: Experimental EPR-steering using Bell-local states}, Nat. Physics {\bf 7}, 918 (2011).
\bibitem{smith}D. H. Smith, G. Gillett, M. P. De Almeida, C. Branciard,
A. Fedrizzi, T. J. Weinhold, A. Lita, B. Calkins, T. Gerrits, H. M. Wiseman, et. al., {\it Conclusive quantum steering with superconducting transition-edge sensors}, Nat. comm. {\bf 3},
625 (2012).

\bibitem{hanchen}V. H\"andchen, T. Eberle, S. Steinlechner, A. Samblowski,
T. Franz, R. F. Werner, and R. Schnabel, {\it Observation of one-way Einstein–Podolsky–Rosen steering}, Nat. Phot. {\bf 6}, 596 (2012).

\bibitem{steinlechner} S. Steinlechner, J. Bauchrowitz, T. Eberle, and R. Schnabel, {\it Strong Einstein-Podolsky-Rosen steering with unconditional entangled states}, Phys. Rev. A {\bf 87}, 022104 (2013).


\bibitem{sun} K. Sun, J.-S. Xu, X.-J. Ye, Y.-C. Wu, J.-L. Chen, C.-F.
Li, and G.-C. Guo, {\it Experimental Demonstration of the Einstein-Podolsky-Rosen Steering Game Based on the All-Versus-Nothing Proof}, Phys. Rev. Lett. {\bf 113}, 140402
(2014).

\bibitem{armstrong} S. Armstrong, M. Wang, R. Y. Teh, Q. Gong, Q. He,
J. Janousek, H.-A. Bachor, M. D. Reid, and P. K. Lam, {\it Multipartite Einstein–Podolsky–Rosen steering and genuine tripartite entanglement with optical networks}, 
Nat. Phys. {\bf 11}, 167 (2015).

\bibitem{cavalcanti} D. Cavalcanti, P. Skrzypczyk, G. Aguilar, R. Nery, P. S.
Ribeiro, and S. Walborn, {\it Detection of entanglement in asymmetric quantum networks and multipartite quantum steering}, Nat. Comm. {\bf 6},
7941 (2015).

\bibitem{cmli} C.-M. Li, K. Chen, Y.-N. Chen, Q. Zhang, Y.-A. Chen,
and J.-W. Pan, {\it Genuine High-Order Einstein-Podolsky-Rosen Steering}, Phys. Rev. Lett. {\bf 115}, 010402
(2015).
\bibitem{deng} X. Deng, Y. Xiang, C. Tian, G. Adesso, Q. He, Q. Gong,
X. Su, C. Xie, and K. Peng, {\it Demonstration of monogamy relations for Einstein-Podolsky-Rosen steering in Gaussian cluster states}, Phys. Rev. Lett.
{\bf 118}, 230501 (2017).
















\end{thebibliography}
\end{document}